\newfam\msbfam
\font\twlmsb=msbm10 at 12pt
\font\eightmsb=msbm10 at 8pt
\font\sixmsb=msbm10 at 6pt
\textfont\msbfam=\twlmsb
\scriptfont\msbfam=\eightmsb
\scriptscriptfont\msbfam=\sixmsb
\def\cj{\fam\msbfam}

\def\C{{\cj C}}

\def\R{{\cj R}}

\def\Z{{\cj Z}}
\def\I{{\cj I}}

\centerline{\bf BUNDLE THEORY OF IMPROPER SPIN TRANSFORMATIONS}

\

\centerline{\bf D. B. Cervantes $^1$, S. L. Quiroga $^2$, L. J. Perissinotti $^{2,3}$ and M. Socolovsky $^1$} 

\

\centerline{\it $^1$ Instituto de Ciencias Nucleares, Universidad Nacional Aut\'onoma de M\'exico}
\centerline{\it Circuito Exterior, Ciudad Universitaria, 04510, M\'exico D. F., M\'exico} 

\centerline{\it $^2$ Departamento de Qu\'\i mica, Facultad de Ciencias Exactas y Naturales,} \centerline{\it Universidad Nacional de Mar del Plata, De\'an Funes 3350, (7600) Mar del Plata, Argentina}
\centerline{\it $^3$ Comisi\'on de Investigaciones Cient\'\i ficas de la Provincia de Buenos Aires, (CIC), Argentina}

\

{\it We first give a geometrical description of the action of the parity operator ($\hat{P}$) on non relativistic spin ${{1}\over{2}}$ Pauli spinors in terms of bundle theory. The relevant bundle, $SU(2)\odot \Z_2\to O(3)$, is a non trivial extension of the universal covering group $SU(2)\to SO(3)$. $\hat{P}$ is the non relativistic limit of the corresponding Dirac matrix operator ${\cal P}=i\gamma_0$ and obeys $\hat{P}^2=-1$. Then, from the direct product of $O(3)$ by $\Z_2$, naturally induced by the structure of the galilean group, we identify, in its double cover, the time reversal operator ($\hat{T}$) acting on spinors, and its product with $\hat{P}$. Both, $\hat{P}$ and $\hat{T}$, generate the group $\Z_4 \times \Z_2$. As in the case of parity, $\hat{T}$ is the non relativistic limit of the corresponding Dirac matrix operator ${\cal T}=\gamma^3 \gamma^1$, and obeys $\hat{T}^2=-1$.}

\


\

{\bf 1. Introduction}

\

As is well known, the effect of proper rotations on non relativistic spin ${{1}\over{2}}$ spinors can be described geometrically through the universal covering group $SU(2)$ of $SO(3)$, that is by the fibre bundle $\xi_s: \Z_2\to SU(2)\buildrel {\pi} \over \longrightarrow SO(3)$. The state vectors are elements of the Hilbert space $\C^2$; in particular , since they are normalized, they belong to the 3-sphere $S^3$ which is topologically equivalent to the group $SU(2)$. The effect of rotations on the ray space can be obtained through the bundle $U(1)\to SU(2)\buildrel {p} \over \longrightarrow \C P^1$ with $p(\pmatrix{z & w \cr -\bar{w} & \bar{z} \cr})=\pmatrix{z \cr w \cr}U(1)$, 
{\it i.e.} by considering the system of bundles (Socolovsky, 2001): $$\matrix{U(1) & & & & \Z_2 \cr & \searrow & & \swarrow & \cr & & SU(2) & & \cr & p\swarrow & & \searrow \pi & \cr \C P^1 & & & & SO(3) \cr}$$ (Topologically, $U(1)\cong S^1$, $\C P^1\cong S^2$, $SO(3)\cong \R P^3$, and $\Z_2 \cong S^0$; in $SU(2)$, $\vert z \vert^2+\vert w \vert ^2=1$.) So, for example, a 2$\pi$ rotation in $\R^3$ leads to a change in the state vector from an initial point on $S^3$ to its antipode; the vector comes back to its initial position only after a 4$\pi$ rotation (Aharonov and Susskind, 1967; Rauch {\it et al}, 1975; Werner {\it et al}, 1975; Silverman, 1980).

On the other hand, it is clear that the effect on spinors of all improper rotations {\it i.e.} those elements in $O(3)$ which are not in its invariant subgroup $SO(3)$, can not be described by $\xi_s$. In particular, we want to find the smallest subgroup of $U(2)$, $S_{\pm}U(2)$ which contains $SU(2)$ and where we can identify the lifting $\hat{P}$ of the {\bf parity transformation} $P=\pmatrix{-1 & 0 & 0 \cr 0 & -1 & 0 \cr 0 & 0 & -1 \cr}$ in $\R ^3$, namely the double cover of $O(3)$, $$\xi:\Z_2\to S_{\pm}U(2)\buildrel {\Pi}\over \longrightarrow O(3). \eqno{(1)}$$ As will be shown below, $S_{\pm}U(2)$ is not simply connected and therefore $S_{\pm}U(2)$ is not the universal covering group of $O(3)$. $SU(2)$ will be an invariant subgroup of $S_{\pm}U(2)$ and $S_{\pm}U(2)$ an invariant subgroup of $U(2)$. $\hat{P}$ in $S_{\pm}U(2)$ will act on spinors $\psi(t,\vec{x})=\pmatrix{u(t,\vec{x}) \cr v(t,\vec{x}) \cr}$ in $SU(2)\cong S^3$ by multiplication: $\psi(t,\vec{x}) \to \psi_P (t,-\vec{x})=
\hat{P} \psi(t,\vec{x})$. Since $\hat{P}$ is the non relativistic limit of the corresponding parity matrix in Dirac theory, which in the standard representation is given by ${\cal P}=\pm i \gamma_0=\pm  i \pmatrix{\I & 0 \cr 0 & -\I \cr}$, where $\I$ is the 2$\times$ 2 unit matrix, then $$\hat{P}=\pm i \I \ \ with \ \ \hat{P}^2=-\I. \eqno{(2)}$$ Then the action on the Pauli spinors is given by $$\hat{P}\psi=\pmatrix{\pm i & 0 \cr 0 & \pm i \cr}\pmatrix{u \cr v \cr}=\pmatrix{\pm iu \cr \pm iv \cr}=\pmatrix{\pm e^{i{{\pi}\over {2}}}u \cr \pm e^{i{{\pi}\over {2}}}v \cr }. \eqno{(3)}$$
As in the relativistic case, two space inversions amount to a 2$\pi$ rotation, which changes the sign of spinors (Berestetskii, Lifshitz and Pitaevskii, 1982; de Azc\'arraga, 1975; Capri, 2002; Socolovsky, 2004). 

\

In addition to the improper rotations, the galilean group of transformations, $G_s$ (de Azc\'arraga and Izquierdo, 1995), consisting of the matrices $$\pmatrix{R & \vec{V} \cr 0 & 1 \cr}, \eqno{(4)}$$ where $R\in SO(3)$ is a rotation, and $\vec{V}\in \R^3$ is the relative velocity between the reference frames, makes natural the consideration of the {\bf time reversal} transformation $T$: we extend $G_s$ to $G$ by replacing $R$ by ${\cal O}\in O(3)$ and 1 by $a \in \{1,-1\}\cong \Z_2$: $$\pmatrix{{\cal O} & \vec{V} \cr 0 & a \cr}. \eqno{(5)}$$ The second row in this matrix precisely gives $t^\prime=-t$ for $a=-1$. If we restrict to $\vec{V}=\vec{0}$ we have the subgroup $G_0$ of $G$ given by $$G_0=\{\pmatrix{{\cal O} & \vec{0} \cr 0 & a \cr}, \ {\cal O}\in O(3), \ a\in \Z_2\} \cong O(3)\times \Z_2. \eqno{(6)}$$ The double cover of this group, $\hat{G}_0$, to be studied in section 4, will contain the three operators $\hat{P}$, $\hat{T}$ and $\hat{P}\hat{T}$ acting on spinors. $\hat{T}$ of course will be the non relativistic limit of the corresponding Dirac matrix operator $\gamma^3\gamma^1$. In sections 2 and 3 we construct $\hat{P}$ and $\xi$, and their corresponding equivalent forms $\hat{P}_\odot$ and $\xi_\odot$.

\

Finally, we want to stress an important point: the fact that even in non relativistic quantum mechanics, $\hat{P}^2=-\I$, leads to the breaking of group isomorphisms between certain double groups (in the old nomenclature). For example, if $\hat{P}^2=\I$ (Landau and Lifshitz, 1997) then the groups  $C_{nv}$, consisting of a principal axis of order $n$ and $n$ reflection planes passing through it, and $D_n$, the dihedric group of order 2$n$, are isomorphic. However, if $\hat{P}^2=-\I$ then these groups become non isomorphic. This result might have experimentally verifiable consequences, in particular in the domain of molecular quantum mechanics, since non isomorphic groups could lead to different spectroscopic selection rules due to the difference in the character tables of their irreducible representations. This controversial subject deserves further research. See, in this connection, Heine (1977) and Sternberg (1997).

\

{\bf 2. $S_{\pm}U(2)$: extension of $SU(2)$ by $\Z_2$}

\

In the relativistic case there is a remaining ambiguity in the sign of ${\cal P}$, and consequently the same occurs in the non relativistic limit. We arbitrarily choose the + sign, $$\hat{P}=i\I=\pmatrix{i & 0 \cr 0 & i \cr} \eqno{(7)}$$ and define the set $$S_{\pm}U(2)=SU(2)\cup SU(2)\hat{P}=SU(2)\cup \{A\hat{P}\}_{A\in SU(2)}. \eqno{(8)}$$ It is clear that $S_{\pm}U(2)$ is a group with $det(A)=1$ and $det(A\hat{P})=-1$. If $A^\prime \in SU(2)$ and $B\in S_{\pm}U(2)$, then $det(BA^\prime B^{-1})=1$ and so $BA^\prime B^{-1} \in SU(2)$ {\it i.e.} $SU(2)$ is an invariant subgroup of $S_{\pm}U(2)$, and if $C \in U(2)$ and $B\in S_{\pm}U(2)$, then $det(CBC^{-1})=det(B)=\pm 1$ and so $CBC^{-1} \in S_{\pm}U(2)$ {\it i.e.} $S_{\pm}U(2)$ is an invariant subgroup of $U(2)$. Being non connected, $S_{\pm}U(2)$ is also non simply connected. 

Let $A=\pmatrix{z & w \cr -\bar{w} & \bar{z} \cr}\in SU(2)$, and $\pi:SU(2)\to SO(3)$ the well known 2$\to$1 homomorphism given by (Naber, 1997) $$\pi(\pmatrix{z & w \cr -\bar{w} & \bar{z} \cr})=\pmatrix{Re(z^2-w^2) & Im(z^2+w^2) & -2Re(zw) \cr -Im(z^2-w^2) & Re(z^2+w^2) & 2Im(zw) \cr 2Re(z\bar{w}) & 2Im(z\bar{w}) & \vert z \vert ^2-\vert w \vert ^2 \cr}. \eqno{(9)}$$ If $\iota$ and $\bar{\iota}$ respectively are the inclusions $SU(2)\buildrel {\iota} \over \hookrightarrow S_{\pm}U(2)$ and $SO(3)\buildrel {\bar{\iota}} \over \hookrightarrow O(3)$, and $\Pi: S_{\pm}U(2)\to O(3)$ is defined by $$\Pi(A)=\pi(A), \ A\in SU(2), \eqno{(10a)}$$ $$\Pi(\hat{P})=\pmatrix{-1 & 0 & 0 \cr 0 & -1 & 0 \cr 0 & 0 & -1 \cr}, \eqno{(10b)}$$ and $$\Pi(A\hat{P})=\Pi(A)\Pi(\hat{P})=-\pi(A), \eqno{(10c)}$$ then: i) the diagram $$\matrix{ SU(2) & \buildrel {\iota} \over \hookrightarrow & S_{\pm}U(2) \cr \pi \downarrow & & \downarrow \Pi \cr SO(3) & \buildrel {\bar{\iota}} \over \hookrightarrow & O(3) \cr} \eqno{(11)}$$ commutes, and ii) $\Pi$ is a group homomorphism (epimorphism). In fact, i) $\Pi(\iota(A))=\Pi(A)=\pi(A)$ and $\bar{\iota}(\pi(A))=\pi(A)$; ii) let $A,A^\prime \in SU(2)$, then $\Pi(AA^\prime)=\pi(AA^\prime)=\pi(A)\pi(A^\prime)=\Pi(A)\Pi(A^\prime)$, $\Pi(A(A^\prime \hat{P}))=\Pi((AA^\prime)\hat{P})=\Pi(AA^\prime)\Pi(\hat{P})=\pi(A)\pi(A^\prime)\Pi(\hat{P})=\pi(A)(\pi(A^\prime)\Pi(\hat{P}))=\Pi(A)\Pi(A^\prime\hat{P})$, $\Pi((A\hat{P})(A^\prime \hat{P}))=\Pi(A\hat{P}^2 A^\prime)=\Pi(A(-\I)A^\prime)=\pi(A)\pi(-\I)\pi(A^\prime)=\pi(A)\pi(A^\prime)$ while $\Pi(A\hat{P})\Pi(A^\prime\hat{P})=\pi(A)\Pi(\hat{P})\pi(A^\prime)\Pi(\hat{P})=\pi(A)\pi(A^\prime)(\Pi(\hat{P}))^2=\pi(A)\pi(A^\prime)$. Moreover, $ker(\pi)=ker(\Pi)=\{\I,-\I\}=\Z_2$.

It can be easily shown that $$1\to SU(2)\buildrel {\iota} \over \hookrightarrow S_{\pm}U(2) \buildrel {det} \over \longrightarrow \Z_2 \to 1 \eqno{(12)}$$ is a short exact sequence of groups and group homomorphisms: $ker(det)=SU(2)=Im(\iota)$ (1 is here the zero group). Then $S_{\pm}U(2)$ is an {\bf extension} of $SU(2)$ by $\Z_2$ (MacLane and Birkoff, 1979). Since $SU(2)$ is non abelian then the extension itself is non abelian and therefore non central. However, (12) {\bf splits}, that is, the map $$\gamma:\Z_2 \to S_{\pm}U(2) \eqno{(13a)}$$ given by $$\gamma(1)=\I, \ and \ \gamma(-1)=\pmatrix{-1 & 0 \cr 0 & 1 \cr}=-\sigma_3 \eqno{(13b)}$$ (one could also define $\gamma(-1)=\sigma_3$) is a group homomorphism which obeys $$det \circ \gamma=Id_{\Z_2} \eqno{(13c)}$$ {\it i.e.} $\gamma$ is a right inverse of $det$. Since $\gamma$ is a monomorphism, then $\Z_2$ is canonically isomorphic to its image $\gamma(\Z_2)=\{\I,-\sigma_3\}$ in $S_{\pm}U(2)$. 

The existence of this splitting allows us to write $S_{\pm}U(2)$ as a semidirect product of $SU(2)$ and $\Z_2$. This is done in the following section. 

\

{\bf 3. $S_{\pm}U(2) \cong SU(2)\odot \Z_2$}

\

{\it Proposition}: Let $\Psi: SU(2)\times \Z_2 \to S_{\pm}U(2)$ be the function defined by $$\Psi(A,B)=AB \eqno{(14)}$$ and $$\phi:\Z_2 \to Aut(SU(2)) \eqno{(15a)}$$ the group homomorphism given by $$\phi(B)(A)=BAB^{-1}. \eqno{(15b)}$$ Then $\Psi$ is a {\bf group isomorphism} with the composition  in $SU(2)\times \Z_2$ given by $$(A^\prime,B^\prime)\cdot (A,B)=(A^\prime \phi(B^\prime)(A),B^\prime B)=(A^\prime B^\prime A {B^\prime}^{-1}, B^\prime B). \eqno{(16)}$$ With this composition law, $SU(2)\times \Z_2$ is the {\bf semidirect product} of $SU(2)$ by $\Z_2$ induced by the action (12b) of $\Z_2$ on $SU(2)$, and is denoted by $SU(2)\odot \Z_2$. 

\

Proof. $\Psi(A,\I)=A\in SU(2)$ and $\Psi(A,-\sigma_3)=\pmatrix{-z & w \cr \bar{w} & \bar{z} \cr}$ with $det(\Psi(A,-\sigma_3))=-1$ {\it i.e.}  

\noindent $\Psi(A,-\sigma_3)\in S_{\pm}U(2)\setminus SU(2)$. Let $\tilde{\Psi}:S_{\pm}U(2)\to  SU(2)\times \Z_2$ be given by $\tilde{\Psi}(A)=(A,\I)$ if $A\in SU(2)$ and $\tilde{\Psi}(B)=(B(-\sigma_3),-\sigma_3)$ if $B\in S_{\pm}U(2)\setminus SU(2)$, then $B=A\hat{P}$ and therefore $B(-\sigma_3)=A\hat{P}(-\sigma_3)=A\pmatrix{-i & 0 \cr 0 & i \cr}=\pmatrix{-iz & iw \cr i\bar{w} & i\bar{z} \cr}\in SU(2)$, and $\Psi(\tilde{\Psi}(B),-\sigma_3)=\pmatrix{iz & iw \cr -i\bar{w} & i\bar{z} \cr}=B$ {\it i.e.} $\tilde{\Psi}=\Psi^{-1}$ and $\Psi$ is a bijection. 

Finally, $\Psi((A^\prime,B^\prime),(A,B))=\Psi(A^\prime B^\prime A {B^\prime}^{-1},B^\prime B)=A^\prime B^\prime A {B^\prime}^{-1}B^\prime B=A^\prime B^\prime AB=\Psi(A^\prime,B^\prime)\Psi(A,B)$ {\it i.e.} $\Psi$ is a group homomorphism.    QED

\

In terms of $SU(2)\odot \Z_2$, one can define the bundle (isomorphic to $\xi$) $$\xi_{\odot}: \Z_2\to SU(2)\odot \Z_2 \buildrel {\Pi_{\odot}} \over \longrightarrow O(3) \eqno{(17)}$$ with $\Pi_{\odot}=\Pi \circ \Psi$ given by $$\Pi_{\odot}(A,B)=\Pi(\Psi(A,B))=\Pi(AB)=\pi(A), \ \ if \ \ B=\I, \eqno{(18a)}$$ and  $$\Pi_\odot(A,B)=\pi(A)\Pi(-\sigma_3)=-\pi(A)\pi(i\sigma_3)=\pi(A)\pmatrix{1 & 0 & 0 \cr 0 & 1 & 0 \cr 0 & 0 & -1 \cr}, \ \ if \ \ B=-\sigma_3. \eqno{(18b)}$$

In $SU(2)\odot \Z_2$, the parity operator is denoted by $\hat{P}_{\odot}$ and is given by  $$\hat{P}_{\odot}=\Psi^{-1}(\hat{P})=\Psi^{-1}(i\I)=(-i\sigma_3,-\sigma_3). \eqno{(19)}$$ This expression is consistent with the commutativity of the diagram $$\matrix{SU(2)\times S^3  & \buildrel {\mu} \over \longrightarrow & S^3 \cr \iota \times Id \downarrow & & \downarrow Id \cr S_{\pm}U(2) \times S^3  & \buildrel {\mu_\iota}\over \longrightarrow & S^3 \cr \Psi^{-1} \times Id \downarrow & & \downarrow Id \cr (SU(2)\odot \Z_2)\times S^3  & \buildrel {\mu_\odot} \over \longrightarrow & S^3 \cr} \eqno{(20)}$$ which gives the maps between the actions $\mu$, $\mu_\iota$ and $\mu_\odot$, respectively of $SU(2)$, $S_{\pm}U(2)$ and $SU(2)\odot \Z_2$ on the Pauli spinors $\psi=\pmatrix{u \cr v \cr}$.

\

{\bf 4. Time reversal $\hat{T}$ and $\hat{P}\hat{T}$}

\

Consider the direct product group $$\hat{G}_0=S_{\pm}U(2)\times \Z_2=(SU(2) \cup  SU(2) \hat{P})\times \Z_2=SU(2)\times \{1\} \cup SU(2)\times \{-1\} \cup \ SU(2)\hat{P}\times \{1\} \cup  SU(2)\hat{P}\times \{-1\}$$ $$=\{(A,1)\}_{A\in SU(2)}\cup \{(A,-1)\}_{A\in SU(2)}\cup \{(A\hat{P},1)\}_{A\in SU(2)}\cup \{(A\hat{P},-1)\}_{A\in SU(2)} \eqno{(21)}$$ and the 2$\to$1 projection $$q:S_{\pm}U(2)\times \Z_2 \to O(3)\times \Z_2 \eqno{(22a)}$$ given by $$q(A,1)=(\pi(A),1), \ q(A,-1)=(\pi(A)R_y(\pi),-1), \ q(A\hat{P},1)=(-\pi(A),1), \ q(A\hat{P},-1)=(-\pi(A)R_y(\pi),-1) \eqno{(22b)}$$ where $$R_y(\pi)=\pmatrix{-1 & 0 & 0 \cr 0 & 1 & 0 \cr 0 & 0 & -1 \cr}=\pi (\pmatrix{0 & -1 \cr 1 & 0 \cr}) \eqno{(22c)}$$ is the $\pi$ rotation around the $y$ axis. Then, in particular, $q(\pmatrix{0 & -1 \cr 1 & 0 \cr},-1)=((R_y(\pi))^2,-1)$

\noindent=$(\pmatrix{1 & 0 & 0 \cr 0 & 1 & 0 \cr 0 & 0 & 1 \cr},-1)$, which gives $t \to t^\prime=-t$ in the spacetime $\R^3\times \R$. It is then natural to identify the {\bf time reversal operator} on spinors as the pair $$(\hat{T},-1)\in \hat{G}_0 \ \ with \ \ \hat{T}=\pmatrix{0 & -1 \cr 1 & 0 \cr}=-i\sigma_2 \in SU(2) \ \ and \ \ \hat{T}^2=-\I, \eqno{(23)}$$ 

\noindent(Feynman, 1987); Sakurai, 1985). A straightforward calculation shows that $q$ is a group homomorphism {\it i.e.} $q((C^\prime,a^\prime)(C,a))=q(C^\prime,a^\prime)q(C,a)$ for all $C^\prime, C$ in $S_{\pm}U(2)$ and $a^\prime, a$ in $\Z_2$. 

\

The four connected components of $\hat{G}_0$ act on spinors as follows: $$(A,1): \ \psi(t,\vec{x})\to \psi_{(A,1)}(t,\vec{x})=(A,1)\cdot \psi(t,\vec{x}):=A\psi(t,\pi(A)\vec{x})=\pmatrix{z & w \cr -\bar{w} & \bar{z} \cr}\pmatrix{u(t,\pi(A)\vec{x}) \cr v(t, \pi(A)\vec{x}) \cr}; \eqno{(24)}$$ 
$$(A,-1): \ \psi(t,\vec{x})\to \psi_{(A,-1)}(t,\vec{x})=(A,-1)\cdot \psi(t,\vec{x}):=A\psi(-t,\vec{x})^*, \eqno{(25)}$$ in particular $$\psi_{(\hat{T},-1)}(t,\vec{x})\equiv \psi_{\hat{T}}(t,\vec{x})=\hat{T}\psi(-t,\vec{x})^*=\pmatrix{0 & -1 \cr 1 & 0 \cr}\pmatrix{u(-t,\vec{x})^* \cr v(-t,\vec{x})^*}=\pmatrix{-v(-t,\vec{x})^* \cr u(-t,\vec{x})^* \cr}, \eqno{(25a)}$$ the time-reversed spinor; $$(B,1): \ \psi(t,\vec{x})\to \psi_{(B,1)}(t,-\vec{x})=(B,1)\cdot \psi(t,\vec{x}):=B\psi(t,\vec{x}), \eqno{(26)}$$ in particular $$\psi_{(\hat{P},1)}(t,-\vec{x})\equiv \psi_{\hat{P}}(t,-\vec{x})=\hat{P}\psi(t,\vec{x})=\pmatrix{i & 0 \cr 0 & i \cr}\pmatrix{u(t,\vec{x}) \cr v(t,\vec{x}) \cr}=\pmatrix{iu(t,\vec{x}) \cr iv(t,\vec{x}) \cr}, \eqno{(26a)}$$ the parity-reversed spinor (3) (with + sign); $$(B,-1): \ \psi(t,\vec{x})\to \psi_{(B,-1)}(t,-\vec{x})=(B,-1)\cdot \psi(t,\vec{x}):=B\hat{T}\psi(-t,\vec{x})^*, \eqno{(27)}$$ in particular $$\psi_{(\hat{P},-1)}(t,-\vec{x})\equiv \psi_{\hat{P}\hat{T}}(t,-\vec{x})=\hat{P}\hat{T}\psi(-t,\vec{x})^*=\sigma_2\psi(-t,\vec{x})^*=\pmatrix{0 & -i \cr i & 0 \cr}\pmatrix{u(-t,\vec{x})^* \cr v(-t,\vec{x})^* \cr}=\pmatrix{-iv(-t,\vec{x})^* \cr iu(-t,\vec{x})^*}, \eqno{(27a)}$$ the time-parity reversed spinor. This is supported by the fact that $$q(\hat{P}\hat{T},-1)=q(i\pmatrix{0 & -1 \cr 1 & 0 \cr},-1)=(-\pi(\pmatrix{0 & -1 \cr 1 & 0 \cr})R_y(\pi),-1)=(-(R_y(\pi))^2,-1)$$ 

\noindent$=(\pmatrix{-1 & 0 & 0 \cr 0 & -1 & 0 \cr 0 & 0 & -1},-1)$ gives $\vec{x}\to \vec{x}^\prime=-\vec{x}$ and $t\to t^\prime=-t$ in $\R^3\times \R$. Clearly $\hat{P}\hat{T}=\hat{T}\hat{P}$ and the matrices $\hat{P}$ and $\hat{T}$ generate the abelian group of order 8 , ${\cal G}_{\hat{P}\hat{T}}$, with multiplication table given by $$\matrix{& \hat{P} & \hat{T} & \hat{P}\hat{T} & -\hat{P} & -\hat{T} & -\hat{P}\hat{T} & -\I \cr \hat{P} & -\I & \hat{P}\hat{T} & -\hat{T} & \I & -\hat{P}\hat{T} & \hat{T} & -\hat{P} \cr \hat{T} & \hat{P}\hat{T} & -\I & -\hat{P} & -\hat{P}\hat{T} & \I & \hat{P} & -\hat{T} \cr \hat{P}\hat{T} & -\hat{T} & -\hat{P} & \I & \hat{T} & \hat{P} & -\I & -\hat{P}\hat{T} \cr -\hat{P} & \I & -\hat{P}\hat{T} & \hat{T} & -\I & \hat{P}\hat{T} & -\hat{T} & \hat{P} \cr -\hat{T} & -\hat{P}\hat{T} & \I & \hat{P} & \hat{P}\hat{T} & -\I & -\hat{P} & \hat{T} \cr -\hat{P}\hat{T} & \hat{T} & \hat{P} & -\I & -\hat{T} & -\hat{P} & \I & \hat{P}\hat{T} \cr -\I & -\hat{P} & -\hat{T} & -\hat{P}\hat{T} & \hat{P} & \hat{T} & \hat{P}\hat{T} & \I \cr} \eqno{(28)}$$ As can be easily verified, ${\cal G}_{\hat{P}\hat{T}}$ is isomorphic to $\Z_4\times \Z_2$ with the isomorphism given by $$\I\to(I,1), \ -\I\to (-I,1), \ \hat{P}\to (\iota,1), \ -\hat{P}\to (-\iota,1),$$ $$\hat{T}\to (\iota,-1), \ -\hat{T}\to (-\iota,-1), \ \hat{P}\hat{T}\to (-I,-1), \ -\hat{P}\hat{T}\to (I,-1), \eqno{(29)}$$ where $\Z_4\cong \{I,\iota,-I,-\iota\}$ with $I$ the identity and $\iota^2=-I$, and $\Z_2\cong \{1,-1\}$. Notice, however, that at the level of spacetime $\R^3\times \R$, $P$ and $T$ generate the group ${\cal G}_{PT}$ with table $$\matrix{& P & T & PT \cr P & 1 & PT & T \cr T & PT & 1 & P \cr PT & T & P & 1 \cr} \eqno{(30)}$$ which, as is well known, is isomorphic to the Klein group $\Z_2\times \Z_2$: $$1\to (1,1), \ P\to (1,-1), \ T\to (-1,1), \ PT\to (-1,-1). \eqno{(31)}$$

\

The group isomorphism (14) between $SU(2)\odot \Z_2$ and $S_{\pm}U(2)$ induces the isomorphism $$\Phi:(SU(2)\odot \Z_2)\times \Z_2 \to S_{\pm}U(2)\times \Z_2, \eqno{(32a)}$$ $$\Phi((A,B),a)=(AB,a), \eqno{(32b)}$$ {\it i.e.} $$\Phi=\Psi \times Id_{\Z_2}. \eqno{(32c)}$$ The group multiplication in the left hand side of (32a) is $$((A^\prime,B^\prime),a^\prime)\cdot ((A,B),a)=((A^\prime B^\prime A {B^\prime}^{-1}, B^\prime B),a^\prime a). \eqno{(33)}$$ Defining $$Q=q\circ \Phi, \eqno{(34)}$$ one has the principal bundle $$\Xi: \Z_2\to (SU(2)\odot \Z_2)\times \Z_2 \buildrel {Q}\over \longrightarrow O(3)\times \Z_2, \eqno{(35)}$$ which generalizes (17). Clearly, $ker(Q)=\Z_2$. {\it This bundle summarizes all the geometry of the parity and time reversal transformations of the Pauli spinors.} 

The inverse of $\Phi$, given by $\Phi^{-1}(A,a)=((A,\I),a)$ for $A\in SU(2)$ and $\Phi^{-1}(B,a)=((-B\sigma_3,-\sigma_3),a)$ for $B\in S_{\pm}U(2)\setminus SU(2)$, gives the operators $\hat{{\cal P}}$ and $\hat{{\cal T}}$ in $(SU(2)\odot \Z_2)\times \Z_2$: $$\hat{{\cal P}}=\Phi^{-1}(\hat{P},1)=((-i\sigma_3,-\sigma_3),1), \eqno{(36a)}$$ and $$\hat{{\cal T}}=\Phi^{-1}=(\hat{T},-1)=((-i\sigma_2,\I),-1). \eqno{(36b)}$$

\

Finally, in the ray space $\C P^1$, the effect of the time reversal, parity and time-parity transformations, is obtained from the definition of the projection $p$ in the introduction, respectively multiplying by $U(1)$ the right hand sides of (25a), (26a) and (27a). 

\

{\bf Acknowledgement}

\

One of us, (M. S.) thanks the hospitality of the Instituto de Astronom\'\i a y F\'\i sica del Espacio (IAFE, UBA-CONICET), Argentina, and of the Departemento de F\'\i sica Te\'orica de la Facultad de Ciencias F\'\i sicas de la Universidad de Valencia, Espa$\tilde{n}$a, where part of this work was performed.

\

{\bf References}

\

Aharonov, Y. and Susskind, L. (1967). Observability of the Sign Change of Spinors under 2$\pi$ Rotations, {\it Physical Review} {\bf 158}, 1237-1238.

\

de Azc\'arraga, J. A. (1975). $P, \ C, \ T, \ \theta$ {\it in Quantum Field Theory}, GIFT 7/75, Zaragoza, Spain: pp. 6-7.

\

de Azc\'arraga, J. A. and Izquierdo, J. M. (1995). {\it Lie groups, Lie algebras, cohomology, and some applications in physics}, Cambridge University Press, Cambridge: p. 153.

\

Berestetskii, V. B., Lifshitz, E. M. and Pitaevskii, L. P. (1982).{\it Quantum Electrodynamics, Landau and Lifshitz Course of Theoretical Physics, Vol. 4}, 2nd. edition, Pergamon Press, Oxford: pp. 69-70.

\

Capri, A. Z. (2002). {\it Relativistic Quantum Mechanics and Introduction to Quantum Field Theory}, World Scientific, New Jersey: pp. 46-51.

\

Feynman, R. P. (1987). The Reason for Antiparticles, in {\it Elementary Particles and the Laws of Physics. 1986 Dirac Memorial Lectures}, eds. R. P. Feynman and S. Weinberg, Cambridge University Press, Princeton, New Jersey: p. 48.

\

Heine, V. (1977). {\it Group Theory in Quantum Mechanics, An Introduction to its Present Usage}, Pergamon Press, Oxford: p. 87.

\

Landau, L. D. and Lifshitz, E. M. (1997). {\it Quantum Mechanics, Non-relativistic Theory, Course of Theoretical Physics, Vol. 3}, 3rd. edition, Butterworth Heinemann, p. 393.

\

MacLane, S. and Birkoff, G. (1979). {\it Algebra}, 2nd. edition, Macmillan Publishing Co., New York: p. 409.

\

Naber, G. L. (1997). {\it Topology, Geometry, and Gauge Fields. Foundations}, Springer-Verlag, New York: pp. 367-377.

\

Rauch, H., Zeilinger, A., Badurek, G. and Wilfing, A. (1975). Verification of coherent spinor rotation of fermions, {\it Physics Letters} {\bf 54A}, 425-427.

\

Sakurai, J. J. (1985). {\it Modern Quantum Mechanics}, Benjamin, Menlo Park, California: p. 278.

\

Silverman, M. P. (1980). The curious problem of spinor rotation, {\it European Journal of Physics} {\bf 1}, 116-123.

\

Socolovsky, M. (2001). On the Geometry of Spin ${{1}\over {2}}$, {\it Advances in Applied Clifford Algebras} {\bf 11}, 487-494.

\

Socolovsky, M. (2004). The CPT Group of the Dirac Field. {\it International Journal of Theoretical Physics} {\bf 43}, 1941-1967; arXiv: math-ph/0404038.

\

Sternberg, S. (1997). {\it Group Theory and Physics}, Cambridge University Press, Cambridge: p. 160.

\

Werner, S. A., Colella. R., Overhauser, A. W. and Eagen, C. F. (1975). Observation of the Phase Shift of a Neutron Due to Precession in a Magnetic Field, {\it Physical Review Letters} {\bf 35}, 1053-1055.

\

\

\

\

\

\

\

e-mails:
daliac@nuclecu.unam.mx, slquirog@mdp.edu.ar, perissin@mdp.edu.ar, 

\noindent socolovs@nuclecu.unam.mx

\end